\documentclass[3p,times,procedia]{elsarticle}
\usepackage{ecrc}
\usepackage{graphics, color, hyperref}
\usepackage{subfig, ulem}
\usepackage{lineno}
\usepackage{multirow}
\usepackage{amssymb}

\newcommand{\1}[1]{~\mathrm{#1}} 

\newcommand{\DBD}[1]{#1$\nu\beta\beta$}

\volume{00}
\firstpage{1}
\journalname{Physics Procedia}
\runauth{The CUORE Collaboration}
 \jid{phpro}
\jnltitlelogo{Physics Procedia}
\CopyrightLine{2014}{Published by Elsevier Ltd.}

\begin{document}

\begin{frontmatter}
\dochead{}

\title{CUORE and beyond: bolometric techniques to explore inverted neutrino mass hierarchy }
\author[USC,LNGS]{D.~R.~Artusa}
\author[USC]{F.~T.~Avignone~III}
\author[INFNLegnaro]{O.~Azzolini}
\author[LNGS]{M.~Balata}
\author[BerkeleyPhys,LBNLNucSci,LNGS]{T.~I.~Banks}
\author[INFNBologna]{G.~Bari}
\author[LBNLMatSci]{J.~Beeman}
\author[Roma,INFNRoma]{F.~Bellini}
\author[INFNGenova]{A.~Bersani}
\author[Milano,INFNMiB]{M.~Biassoni}
\author[Milano,INFNMiB]{C.~Brofferio}
\author[LNGS]{C.~Bucci}
\author[Shanghai]{X.~Z.~Cai}
\author[INFNLegnaro]{A.~Camacho}
\author[LNGS]{L.~Canonica}
\author[Shanghai]{X.~G.~Cao}
\author[Milano,INFNMiB]{S.~Capelli}
\author[INFNMiB]{L.~Carbone}
\author[Roma,INFNRoma]{L.~Cardani}
\author[Milano,INFNMiB]{M.~Carrettoni}
\author[LNGS]{N.~Casali}
\author[Milano,INFNMiB]{D.~Chiesa}
\author[USC]{N.~Chott}
\author[Milano,INFNMiB]{M.~Clemenza}
\author[Genova]{S.~Copello}
\author[Roma,INFNRoma]{C.~Cosmelli}
\author[INFNMiB]{O.~Cremonesi\corref{cor_Cremonesi}}
% ...

\address[USC]{Department of Physics and Astronomy, University of South Carolina, Columbia, SC 29208 - USA}
\address[LNGS]{INFN - Laboratori Nazionali del Gran Sasso, Assergi (L'Aquila) I-67010 - Italy}
\address[INFNLegnaro]{INFN - Laboratori Nazionali di Legnaro, Legnaro (Padova) I-35020 - Italy}
\address[BerkeleyPhys]{Department of Physics, University of California, Berkeley, CA 94720 - USA}
\address[LBNLNucSci]{Nuclear Science Division, Lawrence Berkeley National Laboratory, Berkeley, CA 94720 - USA}
\address[INFNBologna]{INFN - Sezione di Bologna, Bologna I-40127 - Italy}
\address[LBNLMatSci]{Materials Science Division, Lawrence Berkeley National Laboratory, Berkeley, CA 94720 - USA}
\address[Roma]{Dipartimento di Fisica, Sapienza Universit\`a di Roma, Roma I-00185 - Italy}
\address[INFNRoma]{INFN - Sezione di Roma, Roma I-00185 - Italy}
\address[INFNGenova]{INFN - Sezione di Genova, Genova I-16146 - Italy}
\address[Milano]{Dipartimento di Fisica, Universit\`a di Milano-Bicocca, Milano I-20126 - Italy}
\address[INFNMiB]{INFN - Sezione di Milano Bicocca, Milano I-20126 - Italy}
\address[Shanghai]{Shanghai Institute of Applied Physics, Chinese Academy of Sciences, Shanghai 201800 - China}
\address[Genova]{Dipartimento di Fisica, Universit\`a di Genova, Genova I-16146 - Italy}
% ...

\begin{abstract}
The CUORE (Cryogenic Underground Observatory for Rare Events) experiment will search for neutrinoless double beta decay of $^{130}$Te.
With 741 kg of TeO$_2$ crystals and an excellent energy resolution of 5 keV (0.2\%) at the region of interest, CUORE will be one of the most competitive neutrinoless double beta decay experiments on the horizon.
With five years of live time, CUORE projected neutrinoless double beta decay half-life sensitivity is $1.6\times 10^{26}$~y at $1\sigma$ ($9.5\times10^{25}$~y at the 90\% confidence level), which corresponds to an upper limit on the effective Majorana mass in the range 40--100~meV (50--130~meV).
Further background rejection with auxiliary light detector can significantly improve the search sensitivity and competitiveness of bolometric detectors to fully explore the inverted neutrino mass hierarchy with $^{130}$Te and possibly other double beta decay candidate nuclei.
\end{abstract}

\end{frontmatter}
\hrule\vskip12pt
\section*{Abstract}
The CUORE (Cryogenic Underground Observatory for Rare Events) experiment will search for neutrinoless double beta decay of $^{130}$Te.
With 741 kg of TeO$_2$ crystals and an excellent energy resolution of 5 keV (0.2\%) at the region of interest, CUORE will be one of the most competitive neutrinoless double beta decay experiments on the horizon.
With five years of live time, CUORE projected neutrinoless double beta decay half-life sensitivity is $1.6\times 10^{26}$~y at $1\sigma$ ($9.5\times10^{25}$~y at the 90\% confidence level), which corresponds to an upper limit on the effective Majorana mass in the range 40--100~meV (50--130~meV).
Further background rejection with auxiliary light detector can significantly improve the search sensitivity and competitiveness of bolometric detectors to fully explore the inverted neutrino mass hierarchy with $^{130}$Te and possibly other double beta decay candidate nuclei.
\par\vskip10pt
\hrule\vskip12pt

\section{Neutrinoless double beta decay}\label{sec:intro}
The nature of massive neutrinos, i.e, whether neutrinos are their own anti-particles, is of fundamental importance to understand the origin of neutrino mass, and in a broader sense, to explore the symmetries of lepton sector (c.f.~\cite{PDG2012}). 
Neutrinoless double beta decay ($0\nu\beta\beta$) is the only practical probe to the nature of neutrino (c.f.~\cite{Avignone_NDBD_2008, bilenky_NDBD_2012}).

Within the Standard Model, double beta decay with two daughter neutrinos ($2\nu\beta\beta$) 
$(Z,A) \rightarrow (Z+2, A) + 2e^- + 2\bar{\nu}_e$ 
is an allowed 2nd-order weak process and observed in a dozen or so isotopes where the single beta decay is energetically forbidden or kinematically suppressed. 
If neutrinos are their own antiparticles (called Majorana particles),  the $\bar{\nu}_e$ from one single beta decay may undergo helicity flipping and get absorbed in the second beta decay vertex as a $\nu_e$, which results no neutrino emission and  lepton number violation (LNV): 
$(Z,A) \rightarrow (Z+2, A) + 2e^-$.
In fact, $0\nu\beta\beta$ is by far the most realistic mechanism of all the LNV processes~\cite{Vogel_LNV_2013}, which highlights the impact of $0\nu\beta\beta$ beyond neutrino physics.

For \DBD{0} with the exchange of the light Majorana neutrinos, the decay rate $\Gamma$ is:
\begin{equation}\label{eq:decayRate}
\Gamma = G^{0\nu}(Q)|M^{0\nu}|^2\frac{|\langle m_{\beta\beta}\rangle|^2}{m_e^2},
\end{equation}
where , and $G^{0\nu}(Q)$ is phase space factor, $M^{0\nu}$ the nuclear matrix element,  $m_e$ the electron mass, and $|\langle m_{\beta\beta}\rangle|$ the effective Majorana neutrino mass.
$G^{0\nu}(Q)$ is proportional to the 5th order of decay Q-value and can be accurately calculated~\cite{kotila_phase-space_2012}. 
Different model calculations of $M^{0\nu}$ may disagree by a factor of 2 to 3 and introduce large spread of $|\langle m_{\beta\beta}\rangle|$~\cite{Menendez:2008jp,
  PhysRevC.87.014301, Rodriguez:2010mn, Fang:2011da, Faessler:2012ku,
  suhonen_review_2012, Barea:2013bz}.
The effective mass $|\langle m_{\beta\beta}\rangle|$ correlates  $0\nu\beta\beta$ with neutrino mixing parameters closely: $|\langle m_{\beta\beta}\rangle| =\sum_{i=1}^3U_{ei}^2 m_i$, where $U_{ei}$ are the elements in the first row of the neutrino mixing PMNS (Pontecorvo-Maki-Nakagawa-Sakata) matrix~\cite{Pontecorvo_1957, MNS_1962} and $m_i$ the three neutrino mass eigenvalues~\cite{PDG2012}. 
Current generation of \DBD{0} experiments, running or under construction, are generally sensitive to $m_{\beta\beta}$ in the range of 50 to 200~meV, while next generation experiments aim to cover the inverted hierarchy parameter space whose lower boundary is about 10~meV.   

Experimentally, $0\nu\beta\beta$ experiments measure the energy of the two emitting electrons and search for a peak at the decay Q-value in the summed electron energy spectrum.
Currently, more than 10 experiments world-wide aim to search for $0\nu\beta\beta$ utilizing techniques such as ionization, scintillation, thermal excitation, or some combination of those.
For searches for \DBD{0} with half-lives longer than $10^{24}$ years, stringent requirements on parent isotope mass, detector energy resolution, and background rate are all critical.
The commonly accepted figure of merit expression for the half-life limit of $0\nu\beta\beta$  is 
\begin{equation}\label{eq:sens_bkg}
T^{0\nu}_{1/2} \propto \eta\cdot a \cdot \sqrt{\frac{M\cdot t}{b\cdot \Delta E}},
\end{equation}
for background-limited case, where $\eta$ is the physical detector efficiency, $a$ the isotopic abundance of the \DBD{0} candidate, $M$ the total detector mass, $t$ the detector live time, $b$ the background rate per unit detector mass per energy interval, and $\Delta E$ is the energy resolution of the detector~\cite{CUORE_sensitivity_2011}. For the background free (or signal-limited) searches, the half-life limit is proportional to the exposure $M\cdot t$:
\begin{equation}
T^{0\nu}_{1/2}\propto \eta\cdot a\cdot (M\cdot t).
\end{equation}
The detector energy resolution $\Delta E$ does not appear in the equation above, but the ``background-free" criteria impose strict requirement on energy resolution and background rejection of a specific experiment. 

CUORE search for $0\nu\beta\beta$ of $^{130}$Te using a large bolometer array of TeO$_2$ crystals. 
$^{130}$Te is an attractive choice as \DBD{0} emitter for a couple of reasons. The natural isotopic abundance of $^{130}$Te (34.2\%) is highest among all the $0\nu\beta\beta$ candidate isotopes~\cite{Fehr_TeIA_2004}, and therefore no isotopic enrichment is necessary for CUORE. 
The decay Q-value is measured to be around 2528 keV~\cite{Redshaw_Q_2009, Scielzo_Q_2009, Rahaman_Q_2011}, which is higher than most naturally occurring $\gamma$-ray background. 
The relatively high Q-value also means favorable phase space factor in Equation~\ref{eq:decayRate}. 
The previous half-life limit of $^{130}$Te was established by the CUORE predecessor experiment Cuoricino at $2.8\times 10^{24}$ y (90\% C.L.)~\cite{Cuoricino_NDBD_2011}.

CUORE bolometer array features large detector mass and excellent energy resolution at region of interest (ROI).
In this article, we review the advantages of bolometric technique in searching for \DBD{0} and give an update on latest CUORE progress. 
CUORE sensitivity will be limited by background, most of which are $\alpha$ particles from the surface of detector materials and nearby copper thermal shielding. 
Therefore,  future generations of bolometer array, such as the Inverted Hierarchy Explorer (IHE) discussed in~\cite{CUORE_IHE}, focus on background suppression by detecting photon as well as the phonon signal.
The photon signal comes either from Cherenkov radiation or scintillation of \DBD{0} emitting electrons. 
Section~\ref{sec:IHE} discusses the physics and sensitivity in searching for \DBD{0} of $^{130}$Te and other candidate isotopes at the IHE. 

\section{CUORE bolometric technique to search for \DBD{0}}

\subsection{Bolometer design}
\begin{figure}[t!]
 \begin{tabular}{lr}
 \subfloat[]{\label{fig:bolometer}\includegraphics[width=0.4\columnwidth]{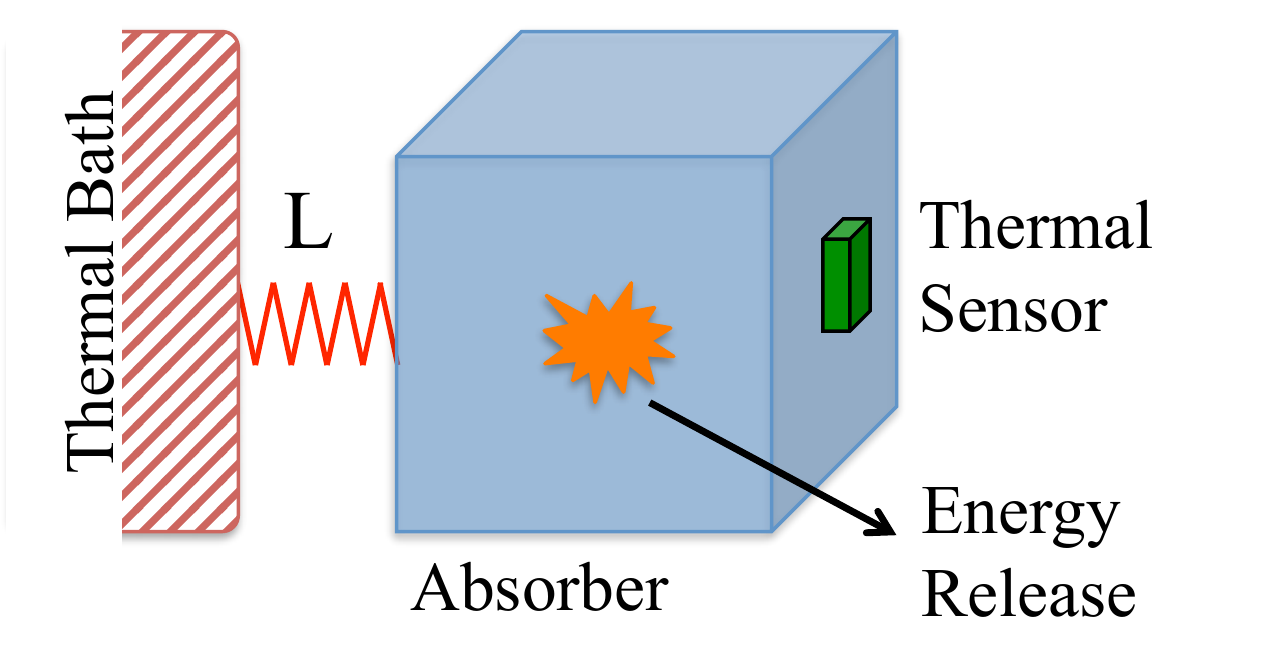}} 
 \hspace{0.1\columnwidth}
 \subfloat[]{\label{fig:scintBolo}{\includegraphics[width=0.4\columnwidth]{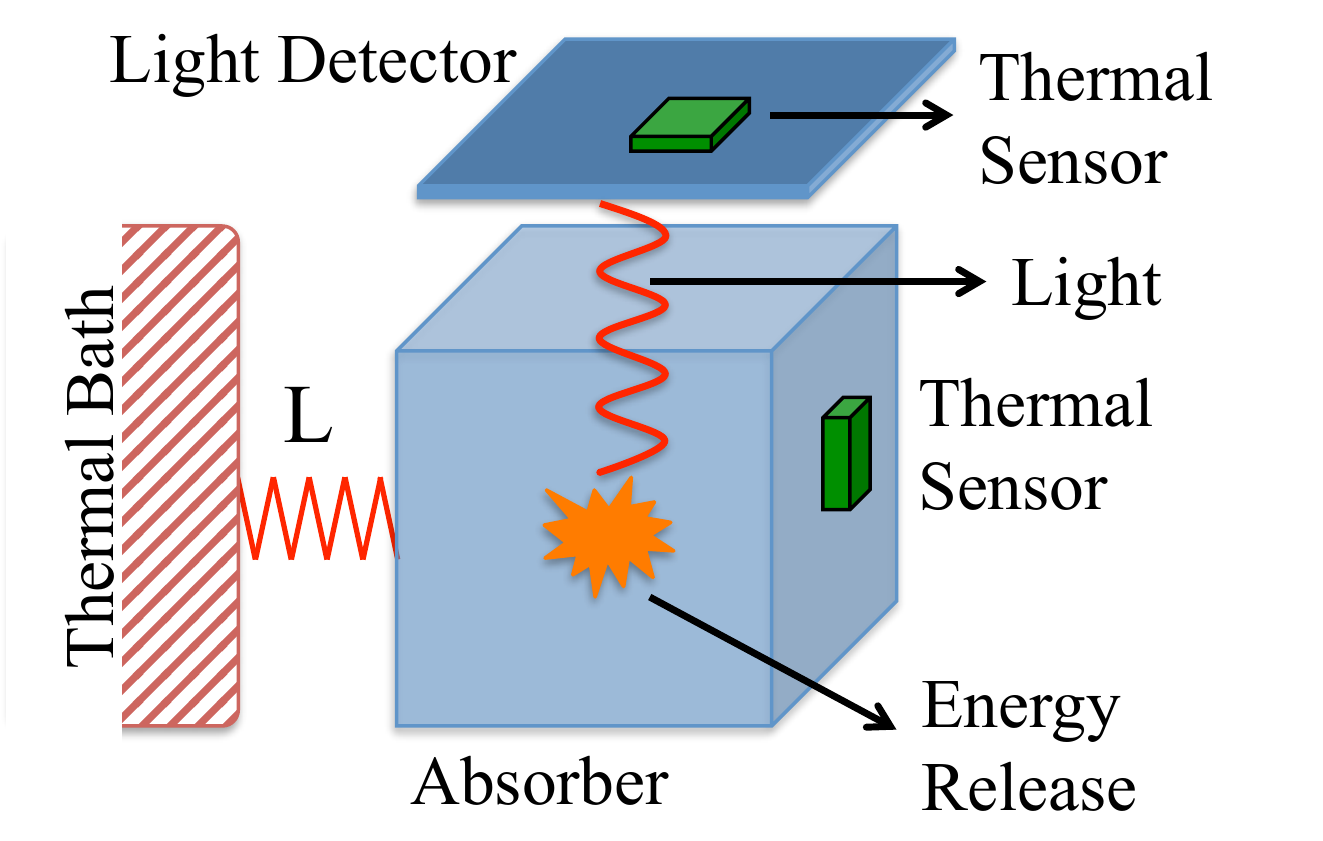}}}\\
 \end{tabular}
\caption{(a)~A schematic setup of typical bolometers. L is weak thermal link between the crystal and the thermal bath. (b)~A similar schematic set for bolometers with an auxillary thin-slab bolometer on top to measure the light output. The light can come from either Cherenkov radiation or scintillation.}
\label{fig:illustration}
\end{figure}

A cryogenic bolometer measures energy release in an absorber via its temperature rise. As shown in Figure~\ref{fig:bolometer}, the temperature rise is usually picked up by a sensitive thermal sensor attached on the absorber.
Because the extremely low phonon excitation energy at cryogenic temperature, the intrinsic sensitivity and resolution of bolometers are unparalleled to any other types of energy detection mechanism.
For example, at 10~mK, the thermal phonon energy is about 1~peV, comparing to $\sim$ 3~eV of an electron-hole pair in semiconductors.   

Bolometers for \DBD{0} searches have massive absorber to maximize the number of the candidate nuclei in it. To minimize the heat capacity of the absorber, dielectric crystals are usually the choice of absorber. 
Thermal sensors are of many different types~\cite{cryogenic_particle_detection}, the most widely used of which are resistive sensors (i.e. thermistors) such as Neutron Transmutation Doped (NTD) germanium thermistor and superconducting Transition Edge Sensor (TES). Microwave Kinetic Inductance Devices (MKIDs) and metallic magnetic sensors have gained popularity in recent years for low temperature detector application. 

\subsection{The CUORE experiment}
The CUORE bolometer array will consist of 988~independent TeO$_2$ crystal modules arranged in 19~towers~(Figure~\ref{fig:towers}).
Each tower has 13 floors with 4 modules on each floor in a two by two configuration.
The main component of each module is a 750~g, 5-cm cubic TeO$_2$ crystal grown from natural tellurium. 
The crystals were manufactured by the Shanghai Institute of Ceramics, Chinese Academy of Sciences with major input from the CUORE collaboration on quality and radio purity control~\cite{Arnaboldi:2010fj}.
The total detector mass is 741~kg and $^{130}$Te isotope mass is 206~kg. 
The NTD Ge thermistor on each module is $3\times3\times1\1{mm^3}$ and  features good reproducibility, excellent stability and low noise. 
The thermistors follow nicely the Mott's law $R=R_0\exp(\sqrt{T_0/T})$~\cite{McCammon:2005yj}, where $T_0\sim4~K$ and $R\sim 1~\Omega$.
Besides the thermal sensor, each bolometer module is also instrumented with a Joule heater, which is used to inject a time-tagged, fixed amount of energy in the bolometer for offline thermal gain correction. 

The Full Width Half Maximum (FWHM) resolution at ROI is expected to be 5~keV. 
Energy resolution of large-mass bolometers are dominated by thermal fluctuation from vibration and readout electronics. 
From CUORE-0 first phase data, we evaluate the overall detector energy resolution to be 5.7~keV, based on the FWHM of the 2615~keV peak in the energy spectrum for \DBD{0} searches~\cite{CUORE0_InitialPerformance_2014, TAUP2013_CUORE-0}. 
Recently, we have reduced the vibration from the CUORE-0 cryostat and saw noticeable improvement in the energy resolution. 
It should be also noted that CUORE-0 has been running at a base temperature range 13--15~mK, higher than expected CUORE base temperature at 10~mK.
In our R\&D runs, when the bolometers were operated below 13~mK, we consistently achieved energy resolution below 5~keV~\cite{CUORE_AHEP_2014}. 

The projected background rate at ROI is $0.01\1{counts/keV/kg/y}$, which is mainly from surface and bulk contaminations of the bolometer components and the innermost copper thermal shield.
Surface-related background, mainly $\alpha$ particles originated from the copper surface is 1--2$\times10^{-2}$~counts/keV/kg/y, based on Monte Carlo simulations with the latest results of CUORE-0~\cite{CUORE_projected_BG}. 
Background contribution from bulk contamination, predominately $\gamma$ events from $^{208}$Tl, is expected to be less than $6\times10^{-3}$~counts/keV/kg/y~(90\% C.L.)~\cite{CUORE_projected_BG}.

CUORE will be situated at the LNGS underground facility at an average depth of 3650~m water equivalent, where the muon flux is $(2.58\pm0.3)\times10^{-8}\1{\mu/(s\cdot cm^2)}$, about six order of magnitude smaller than that of sea level~\cite{Mei:2005gm}. 
The CUORE cryostat will be surrounded by borated polyethylene, boric-acid powder, and lead bricks to attenuate neutron and $\gamma$-ray backgrounds.
More lead shielding is added inside the cryostat, including ancient Roman lead~\cite{Alessandrello:1998xy} to further suppress the $\gamma$-rays.
We find the expected environmental muon, neutron, and $\gamma$ background rates in the ROI to be orders of magnitude smaller than the $\alpha$ and $\gamma$ backgrounds from the experimental apparatus itself~\cite{Andreotti:2009dk, CUOREExternal}.

 CUORE sensitivity is limited by the background and follows Equation~\ref{eq:sens_bkg}. A more rigorous sensitivity projection based on Poissonian background fluctuation shows that CUORE projected half-life sensitivity is $1.6\times 10^{26}$~y at $1\sigma$ ($9.5\times10^{25}$~y at the 90\% C.L.)~\cite{CUORE_sensitivity_2011}, which corresponds to an upper limit on the effective Majorana mass in the range 40 --100 meV (50 --130 meV). The  lower and upper bound of $m_{\beta\beta}$ is calculated with the most and least favorable NME~\cite{Menendez:2008jp,
  PhysRevC.87.014301, Rodriguez:2010mn, Fang:2011da, Faessler:2012ku,
  suhonen_review_2012, Barea:2013bz}.

\subsection{Additional advantage of bolometer for \DBD{0} search}
Typically we categorize searches for \DBD{0} into two groups by the relation of \DBD{0} source and detector. One common approach is to implement a detector with candidate isotope  inside.  For example, GERDA uses semiconductor $^{76}$Ge-enriched germanium diode to search for \DBD{0} of $^{76}$Ge~\cite{GERDA_2013}. The ``source in detector" approach has the advantage of high detecting efficiency. The other approach emphasizes on the flexibility of \DBD{0} source choices by separating the source from detection, the typical example of which is the NEMO3/SuperNEMO project~\cite{barabash_supernemo_2011}. 

Bolometric technique combines the benefits of high efficiency and flexibility of candidate isotope choice. For large-mass bolometers such as CUORE modules, the probability of both \DBD{0} electrons are confined in the crystal is $(87.4\pm 1.1)\%$. Bolometric technique puts little constraints on the absorber materials, other than preferably dielectric for the benefit of small heat capacity. Crystals such as TeO$_2$ or ZnSe are easily interchangeable for search of \DBD{0} in different candidate isotopes. The flexibility is especially useful for confirming \DBD{0} discoveries or cross-comparing $m_{\beta\beta}$ limits in different isotopes.

\section{CUORE status}
\label{sec:status}

\begin{figure}[t!]
 \begin{tabular}{lr}
 \subfloat[]{\label{fig:towers}\includegraphics[width=0.5\columnwidth]{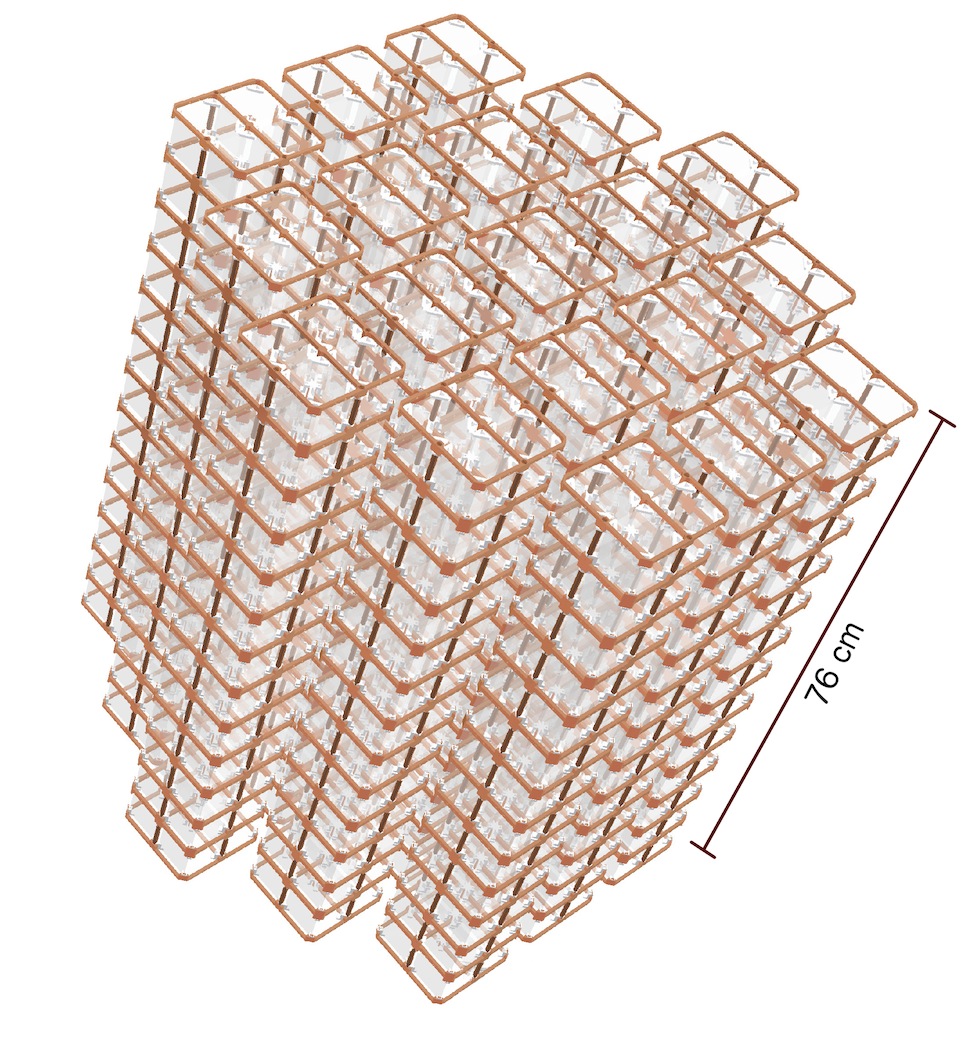}} 
 \hspace{0.04\columnwidth}
 \subfloat[]{\label{fig:cryostat}{\includegraphics[width=0.4\columnwidth]{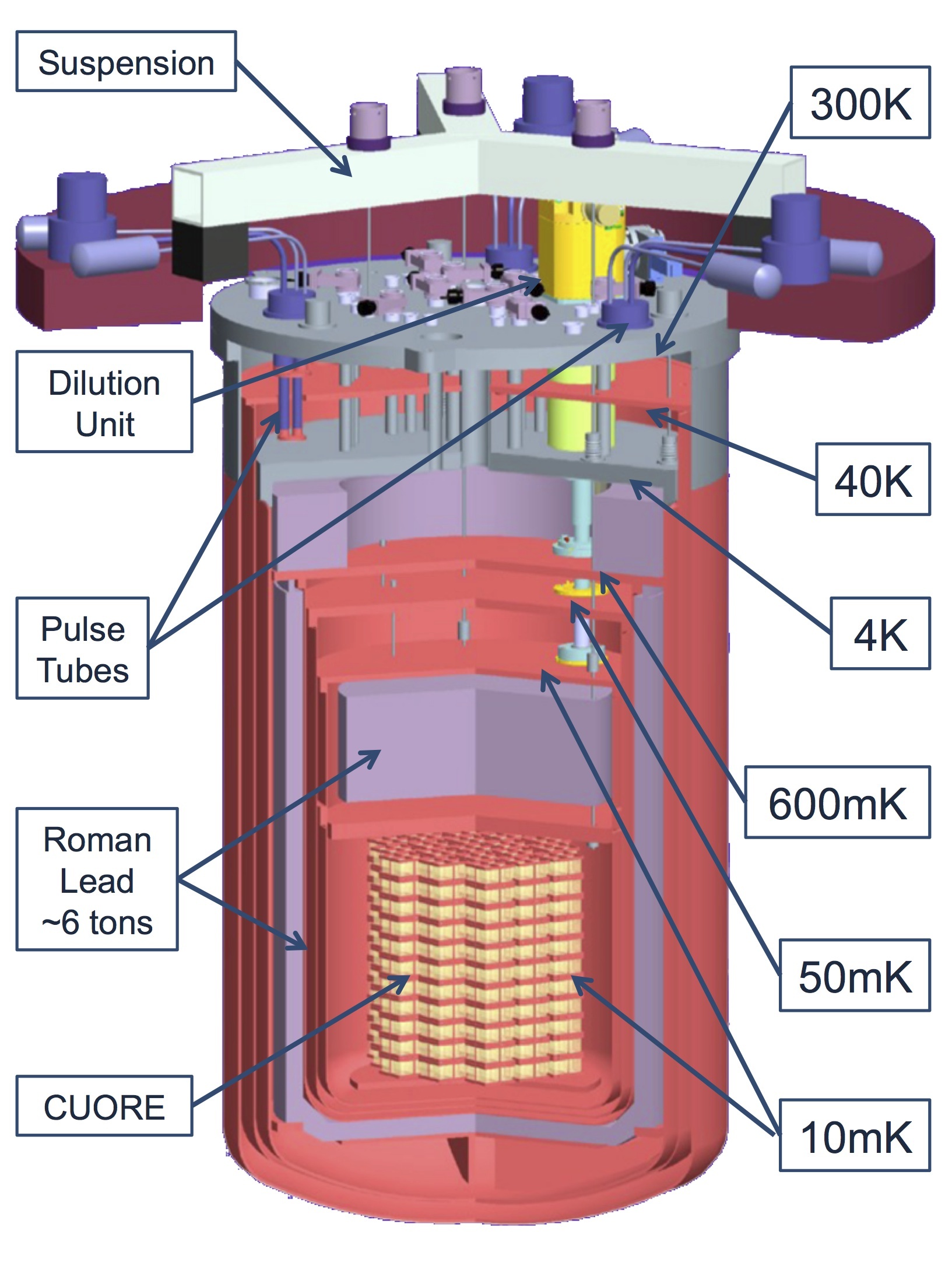}}}\\
 \end{tabular}
\caption{(a)~Schematic setup of CUORE towers. 19 towers are closely packed. Each tower has 13 floors and 4 bolometer modules on each floor. (9) CUORE cryostat with major components highlighted.}
\label{fig:apparatus}
\end{figure}
\begin{figure}[t]
 \begin{tabular}{lr}
 \subfloat[]{\label{fig:half-life}\includegraphics[height=0.4\columnwidth]{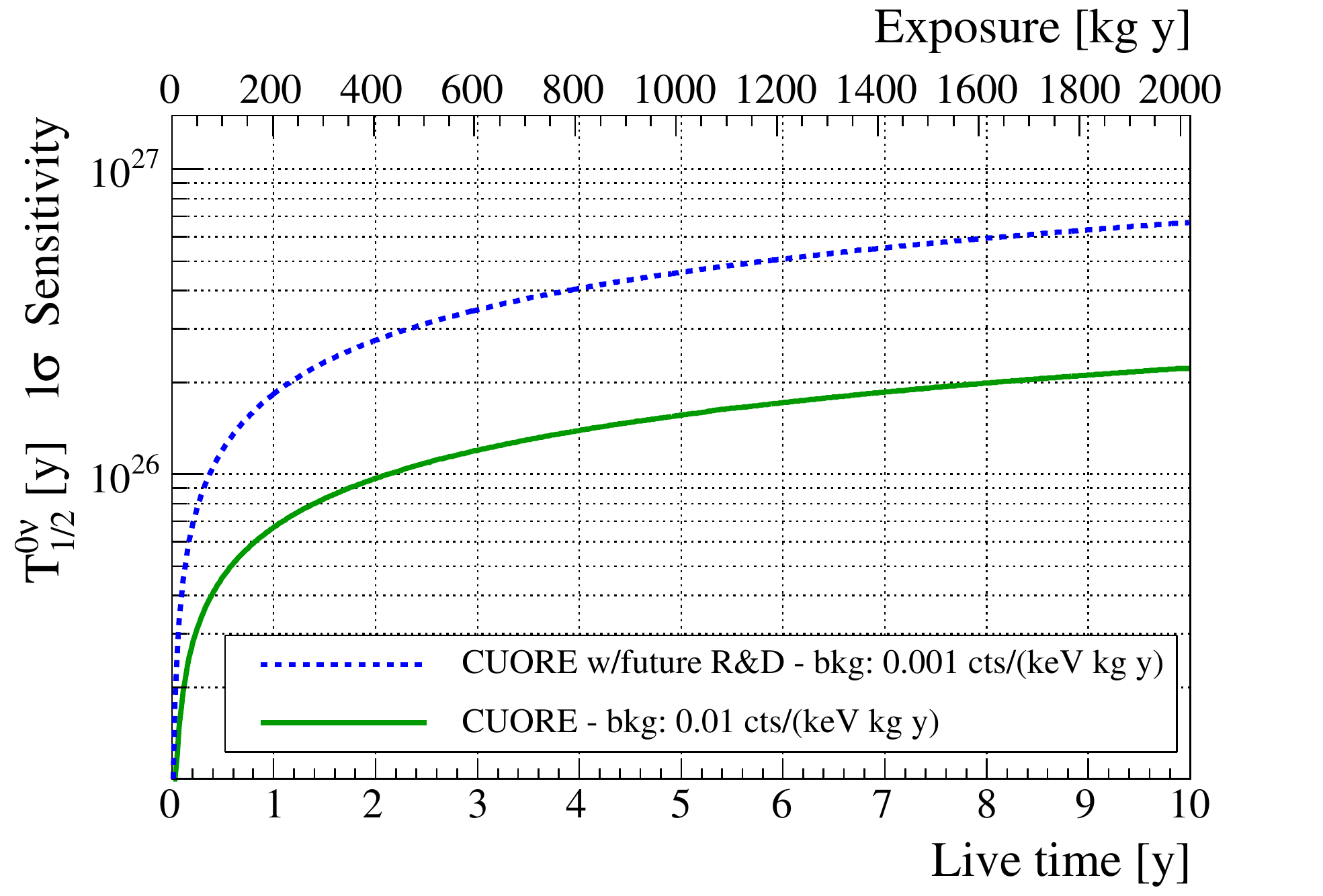}} 
\subfloat[]{\label{fig:mee}\includegraphics[height=0.39\columnwidth]{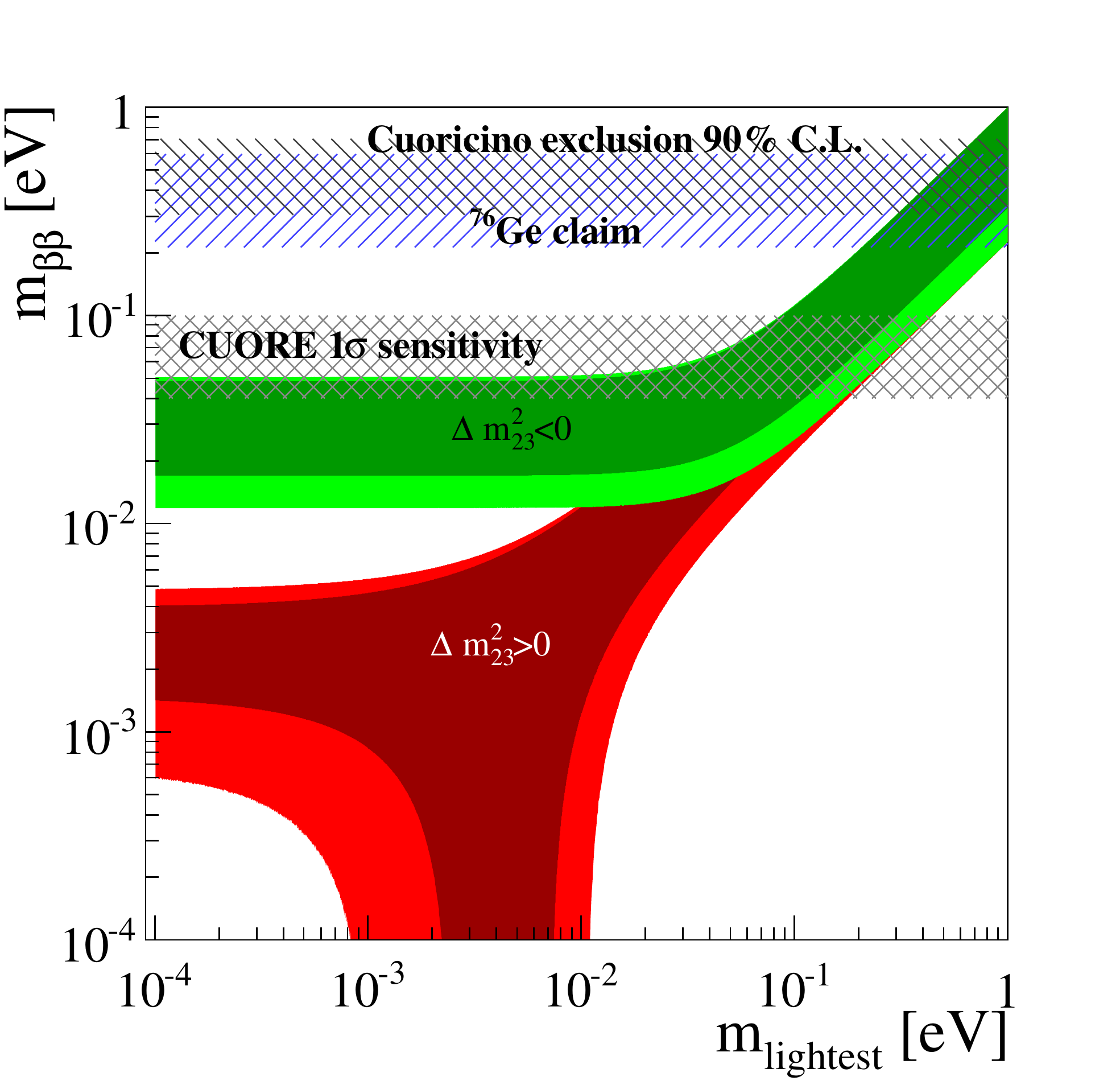}}\\
 \end{tabular}
\caption{(a)~Projected CUORE $1\sigma$ sensitivity to the half-life of \DBD{0} of $^{130}$Te as a function of detector live time. The solid green line shows the sensitivity for the target background rate of 0.01~counts/keV/kg/y, while the dashed blue line shows an optimistic scenario of 0.001~counts/keV/kg/y background rate. (b)~The corresponding $1\sigma$ sensitivity to effective Majorana mass versus lightest neutrino mass after five years of detector live time. The spread in the projected CUORE $m_{\beta\beta}$ bands arises from uncertainties in calculations of the nuclear matrix elements used to convert a measured half-life into an effective Majorana mass. The green band labeled $\Delta m^2_{23}<0$ denotes the inverted neutrino mass hierarchy while the red band labeled $\Delta m^2_{23}>0$ indicates the normal mass hierarchy. For both, the darker inner bands represent regions allowed by the best-fit neutrino oscillation parameters in the PMNS matrix, while the lighter outer bands extend to $3\sigma$ coverage. Both figures are from~\cite{CUORE_sensitivity_2011}.}
\end{figure}

Ongoing CUORE construction includes major effort on bolometer array assembly, cryostat commissioning, detector calibration system (DCS) integration, electronics and software development.
\subsection{Detector assembly}
\label{sec:assembly}
During the bolometer array assembly process, we need to turn 10000 loose parts into 988 ultra-clean bolometer modules. 
All the operations are conducted in glove boxes under nitrogen atmosphere to minimize oxidization and contamination of radon and radon progenies~\cite{Clemenza:2011zz}.
The assembly process  starts by gluing of thermistors and heaters to crystals with a robotic system to maintain the uniformity and repeatability of the gluing joints. 
Then we assemble the instrumented crystals, cleaned copper parts, and PTFE spacers into a tower. 
Along the two wire trays on opposite side of each tower, we attach  two sets of flexible printed circuit board readout cables, whose top ends will be connected to the cryostat wiring later.
The bottom ends of the readout cables are terminated with copper bonding pads.
In the final step of detector assembly, we use a wire bonder to bond the thermistors and heaters to the bonding pads to complete the electrical connections.

CUORE tower assembly began in January 2013 and will be finished by summer of 2014. 
As of this writing we have glued 15~tower's worth of crystals, physically assembled 13~towers, and wire bonded 12.
More details on CUORE detector assembly can be found in~\cite{CUORE_AHEP_2014}.

\subsection{Cryogenics}

CUORE cryogenics system cools the bolometer array and the immediate shielding,  almost 7 ton of material, to 10~mK base temperature.
The system includes a large cryostat with six nested thermal shields, a cryogen-free cooling system with five pulse-tube coolers, an auxiliary fast-cooling system for pre-cooling, and a dilution-refrigerator unit~\cite{Cryostat_overview_2012}.

All six copper thermal shields of the cryostat have been delivered to LNGS underground facility for commissioning.
In the first phase of cryostat commissioning, we cooled down the outer three vessels with three pulse tubes twice, successfully reaching 3.5~K on the 4~K plate on the second attempt~\cite{Cryostat_4K_2014}.
The second phase of commissioning with three inner vessels is currently ongoing.
In addition to the pulse tubes and closed-loop pre-cooling circulation lines, an auxiliary fast-cooling system circulates helium gas through the cryostat vessels to speed up the cooling from room temperature to 4~K. 

The custom-built $^{3}$He/$^{4}$He dilution unit by Leiden Cryogenics was delivered in Summer 2012 after passing in-house benchmarking. 
At LNGS it was tested in a small testing cryostat and successfully reached 5~mK base temperature with a cooling power of $5\1{\mu W}$ at 12~mK.  The unit will be integrated to the main cryostat after the commissioning of the inner thermal shields.

\subsection{Detector Calibration System} 
DCS is designed to insert and retrieve 12~radioactive source strings from the 300~K top flange to the bolometer array at 10~mK for energy calibrations every month.
The main challenge is to limit the heat-load while lowering the source strings through the successive flanges with decreasing temperatures.
Along each Kevlar string, we crimp 25 thoriated tungsten sources, evenly separated by 29 mm from each other. 
A motion control system on top of the 300~K flange releases each string to travel down a dedicated guide tube by its own weight.
The strings are thermalized at the 4~K flange to avoid over-heating the three inner cryostat vessels. 
DCS was successfully tested to the 4~K flange together with a commissioning cooldown of the outer cryostat as mentioned earlier. 
We will further test the DCS along with future cryostat commissioning cool downs.

\subsection{Electronics, data acquisition, and analysis}
The CUORE electronics will read the thermistor voltage signal to computer via the front end electronics, high-precision 18-bit digitizers, and the data acquisition software \textsc{Apollo}.
Each front-end channel consists of load resistors, a detector biasing system, a two-stage amplifier system with programmable gain up to 10000, and a 6-pole anti-aliasing Bessel filter with cutoff frequencies in the range 15--120~Hz.
We have prepared a single chassis system with a full chain CUORE electronics for testing on a mini bolometer array during upcoming commissioning tests of the CUORE cryostat.

The DAQ software \textsc{Apollo} reads in the digitized waveforms at 125~S/s, runs trigger algorithms, and records both the continuous and triggered data on the disk for offline analysis.
\textsc{Apollo} has been used successfully for CUORE-0 data taking~\cite{CUORE0_InitialPerformance_2014}. Current development focuses on scaling from one DAQ computer use case (as in CUORE-0) to a small DAQ cluster to acquire the $\sim$1000 channels in CUORE.
Our custom-built data analysis software framework \textsc{Diana} has been used extensively for analysis of Cuoricino, R\&D runs, and more recently  CUORE-0.
New features such as noise decorrelation~\cite{ManciniTerracciano:2012fq} and browser-based data quality monitoring are being implemented.

\section{Inverted Hierarchy Explorer -- bolometer with light output}
\label{sec:IHE}

\begin{table*}
\begin{center}
\begin{tabular}{cccccccccc}
\hline\hline
Isotope & Q-Value  & Crystal 			& Exposure  & Background &$|m_{\beta\beta}|_{discovery}$	 & $\widehat{T^{0\nu}_{1/2}}$  & $|m_{\beta\beta}|_S$ \\
 &[keV]    &  &      [ton$\cdot$y] & [counts/ton/y]  & [meV]         	 &         [10$^{27}$y]   & [meV]  \\
\hline
 $^{82}$Se	& 2996 	  & ZnSe  				&	    3.3       &     0.1  & 18 - 52		& 2.2    & 11 - 32 \\
$^{116}$Cd	& 2814 		&  CdWO$_4$	 		&	    4.9  &       0.1     & 24 - 45		& 1.5    &14 - 26 \\
$^{100}$Mo	& 3034		&ZnMoO$_4$ 			&	    2.7           &   1.5 & 24 - 69		&  0.65 & 13 - 37 \\
$^{130}$Te	& 2528 	 	& TeO$_2$  			&	    3.7            &  0.1 & 17 - 43 		& 2.6    & 10 - 25 \\
\hline\hline
\end{tabular}
\end{center}
\caption{Four candidate isotopes considered for the IHE. The exposure assumes five year live time. Background rates are assumed to be 0.1 counts/ton/y except for the $^{100}$Mo case, which is higher because of \DBD{2} events pile-up. Assuming neutrinos are of Majorana type, IHE would discover a signal of $5\sigma$ or larger if the effective Majorana mass is at least $|m_{\beta\beta}|_{discovery}$.  $\widehat{T^{0\nu}_{1/2}}$  and $|m_{\beta\beta}|_S$ are 90\% C.L. sensitivity to \DBD{0} half-life and effective Majorana mass. The range of $|m_{\beta\beta}|$ reflects the uncertainties of different model calculations of NME. Data from~\cite{CUORE_IHE}.}
\label{tab:resultsIHE}
\end{table*}

Background suppression with particle identification is the main emphasis on bolometer's future application in rare event searches. 
Additional detection channels are needed for particle identification since the absorber does not respond differently for energy release of different particle types. 
To distinguish between \DBD{0} electrons and $\alpha$ particle background, light emission, either from  Cherenkov radiation or scintillation, can be exploited.
The auxiliary light detector is usually another bolometer facing the main bolometer.
The light detector nominally consists of a thin germanium/silicon wafer as absorber and a thermal sensor of the same type as the main bolometer.  

The next generation bolometric experiment with light output, which we call Inverted Hierarchy Explorer (IHE), aims to explore the inverted neutrino mass hierarchy region at effective Majorana mass $m_{\beta\beta}\sim 10\1{meV}$~\cite{CUORE_IHE}.  We have studied the sensitivities of TeO$_2$ bolometers with Cherenkov light readout and three scintillating bolometers ZnSe,  CdWO$_4$, and ZnMoO$_4$, all of which are hosted in a dilution refrigerator similar to CUORE (see Figure~\ref{fig:cryostat}). 

The TeO$_2$ crystal is largely transparent to Cherenkov light radiation generated by electrons with energy larger than 50 keV~\cite{TabarellideFatis:2009zz}. Should the Cherenkov light from electrons observable, it provides a definite discrimination between $\alpha$ backgrounds and electrons. For typical bolometer $\alpha$ background, their energy is an order of magnitude lower than the Cherenkov radiation threshold in TeO$_2$. Cherenkov light emission in TeO$_2$ has been detected using a secondary light detector with modest particle identification capability~\cite{Beeman:2011yc}. We are actively developing better light collection schemes and more importantly, light detectors with lower threshold and better energy resolution. Future directions include new thermal sensors on the light detector, such as TES~\cite{Angloher:2011uu} and MKIDs~\cite{Calder_JLTP_2014},  and Luke effect~\cite{Isaila2012160} enhanced bolometers.

A scintillating bolometer operates similarly to the Cherenkov case, but with a different crystals and more light output from scintillation. The particle identification comes from the different light yield for different particle type.
For scintillating crystals such as ZnSe, CdWO$_4$, and ZnMoO$_4$, the light yield difference between possible \DBD{0} electrons and $\alpha$ particles are well measured~\cite{Pirr06,Arnaboldi:2010jx,Beeman2013JINST,Arnaboldi:2010tt,Beeman2012318,Gironi:2010hs,ZnMoO4_1,ZnMoO4_2}.

IHE detector will consist of 988 bolometer modules, each with a 5-cm cubic crystal of our choice.  The detector arrangement follows CUORE design. The four bolometer modules on each floor shares a light detector made of radio-pure germanium or silicon wafer of about 300~$\mu$m thick and 10~cm in diameter. The crystals are wrapped in reflecting foils to improve the light collection efficiency. 
The natural isotopic abundance of $^{82}$Se, $^{116}$Cd and $^{100}$Mo are all less than 10\%. In IHE, we assume all \DBD{0} candidate isotopes are enriched to 90\%. 

The physics reach of IHE for the different \DBD{0} candidate isotopes are reported in Table~\ref{tab:resultsIHE} with 5 years of live time and $\Delta E$= 5~keV. In the table, we make  aggressive assumption on background reduction and an optimistic background index of 0.1~counts/ton/y (1.5 counts/ton/y for the ZnMoO$_4$  case due to the pile-up of the intrinsic \DBD{2} events) is assumed. More information about the background budget and effectiveness of background suppression can be found in~\cite{CUORE_IHE}. 
In the last three columns of the table, we present the minimum observable value of the effective Majorana mass $|m_{\beta\beta}|_{discovery}$ for a \DBD{0} discovery and 90\% C.L. sensitivity on half-life and $|m_{\beta\beta}|_S$ for a null result.
Assuming neutrinos are of Majorana type, IHE would discover a signal of $5\sigma$ or larger if the effective Majorana mass is at least $|m_{\beta\beta}|_{discovery}$. 
The 90\% C.L. half-life sensitivities are all on the $10^{27}$ y level. The TeO$_2$ case is especially sensitive to the effective Majorana mass at 10 to 25 meV, reaching the lower boundary of the inverted mass hierarchy. 

\section{Conclusion and outlook}
We reviewed the physics reach and current construction status of the CUORE experiment. Data taking is expected to start in 2015. With excellent energy resolution and large isotope mass, CUORE is one of the most competitive \DBD{0} experiments under construction. 

Future bolometer program such as the IHE can explore the inverted mass hierarchy region to $|m_{\beta\beta}|\sim 10\1{meV}$ by utilizing additional light readout to further reduce background. We described the conceptual design of IHE and its physics impact on $^{130}$Te, $^{82}$Se, $^{116}$Cd or $^{100}$Mo \DBD{0} searches.
\section*{Acknowledgements}

The CUORE Collaboration thanks the directors and staff of the Laboratori Nazionali del Gran Sasso and the technical staff of our laboratories. 
This work was supported by the Istituto Nazionale di Fisica Nucleare (INFN); the Director, Office of Science, of the U.S. Department of Energy under Contract Nos. DE-AC02-05CH11231 and DE-AC52-07NA27344; the DOE Office of Nuclear Physics under Contract Nos. DE-FG02-08ER41551 and DEFG03-00ER41138; the National Science Foundation under Grant Nos. NSF-PHY-0605119, NSF-PHY-0500337, NSF-PHY-0855314, NSF-PHY-0902171, and NSF-PHY-0969852; the Alfred P. Sloan Foundation; the University of Wisconsin Foundation; and Yale University. 
This research used resources of the National Energy Research Scientific Computing Center (NERSC).

\bibliographystyle{elsarticle-num}
\bibliography{CUORE_and_Beyond_KeHan2013}

\begin{thebibliography}{10}
\expandafter\ifx\csname url\endcsname\relax
  \def\url#1{\texttt{#1}}\fi
\expandafter\ifx\csname urlprefix\endcsname\relax\def\urlprefix{URL }\fi
\expandafter\ifx\csname href\endcsname\relax
  \def\href#1#2{#2} \def\path#1{#1}\fi

\bibitem{PDG2012}
J.~Beringer, et~al., The review of particle physics, Phys. Rev. D86 (2012)
  010001.

\bibitem{Avignone_NDBD_2008}
F.~T. Avignone, S.~R. Elliott, J.~Engel, {Double beta decay, Majorana
  neutrinos, and neutrino mass}, Rev. Mod. Phys. 80~(2) (2008) 481--516.

\bibitem{bilenky_NDBD_2012}
S.~M. Bilenky, C.~Giunti, Neutrinoless double-beta decay. a brief review,
  {arXiv:1203.5250}.

\bibitem{Vogel_LNV_2013}
A.~de~Gouvea, P.~Vogel, Lepton flavor and number conservation, and physics
  beyond the standard model, {arXiv:1303.4097}.

\bibitem{kotila_phase-space_2012}
J.~Kotila, F.~Iachello, Phase-space factors for double-β decay, Physical
  Review C 85~(3) (2012) 034316.
\newblock

\bibitem{Menendez:2008jp}
J.~Menendez, A.~Poves, E.~Caurier, F.~Nowacki, {Disassembling the Nuclear
  Matrix Elements of the Neutrinoless beta beta Decay}, Nucl. Phys. A 818
  (2009) 139--151.
\newblock \href {http://arxiv.org/abs/0801.3760} {\path{arXiv:0801.3760}},

\bibitem{PhysRevC.87.014301}
P.~K. Rath, et~al., Uncertainties in nuclear transition matrix elements for
  $\beta+\beta+$ and $\epsilon\beta$+ modes of neutrinoless positron
  double-$\beta$ decay within the projected hartree-fock-bogoliubov model,
  Phys. Rev. C 87 (2013) 014301.
\newblock

\bibitem{Rodriguez:2010mn}
T.~R. Rodriguez, G.~Martinez-Pinedo, {Energy density functional study of
  nuclear matrix elements for neutrinoless $\beta\beta$ decay}, Phys. Rev.
  Lett. 105 (2010) 252503.
\newblock \href {http://arxiv.org/abs/1008.5260} {\path{arXiv:1008.5260}},

\bibitem{Fang:2011da}
D.-L. Fang, A.~Faessler, V.~Rodin, F.~Simkovic, {Neutrinoless double beta decay
  of deformed nuclei within QRPA with realistic interaction}, Phys. Rev. C 83
  (2011) 034320.
\newblock \href {http://arxiv.org/abs/1101.2149} {\path{arXiv:1101.2149}},

\bibitem{Faessler:2012ku}
A.~Faessler, V.~Rodin, F.~Simkovic, {Nuclear matrix elements for neutrinoless
  double-beta decay and double-electron capture}, J. Phys. G: Nucl. Part. Phys.
  39 (2012) 124006.
\newblock \href {http://arxiv.org/abs/1206.0464} {\path{arXiv:1206.0464}},

\bibitem{suhonen_review_2012}
J.~Suhonen, O.~Civitarese, Review of the properties of the $0\nu\beta\beta$
  nuclear matrix elements, J. Phys. G: Nucl. Part. Phys. 39~(12) (2012) 124005.

\bibitem{Barea:2013bz}
J.~Barea, J.~Kotila, F.~Iachello, {Nuclear matrix elements for double-{$\beta$}
  decay}, Phys. Rev. C 87 (2013) 014315.
\newblock \href {http://arxiv.org/abs/1301.4203} {\path{arXiv:1301.4203}},

\bibitem{Pontecorvo_1957}
B.~M. Pontecorvo, Mesonium and antimesonium, J. Exp. Theor. Phys. Lett. 33
  (1957) 549.

\bibitem{MNS_1962}
Z.~Maki, M.~Nakagawa, S.~Sakata, Remarks on the unified model of elementary
  particles, Prog. Theor. Phys. 28~(5) (1962) 870--880.

\bibitem{CUORE_sensitivity_2011}
F.~Alessandria, et~al., Sensitivity of {CUORE} to neutrinoless double-beta
  decay, arXiv:1109.0494.

\bibitem{Fehr_TeIA_2004}
M.~A. Fehr, M.~Rehkamper, A.~N. Halliday, Application of {MC-ICPMS} to the
  precise determination of tellurium isotope compositions in chondrites, iron
  meteorites and sulfides, Int. J. Mass spectrom. 232 (2004) 83--94.

\bibitem{Redshaw_Q_2009}
M.~Redshaw, B.~J. Mount, E.~G. Myers, F.~T. Avignone, III, {Masses of
  $^{130}$Te and $^{130}$Xe and double-$\beta$-decay Q-value of $^{130}$Te},
  Phys. Rev. Lett. 102~(21) (2009) 212502.
\newblock \href {http://arxiv.org/abs/0902.2139} {\path{arXiv:0902.2139}}.

\bibitem{Scielzo_Q_2009}
N.~D. Scielzo, et~al., {Double-$\beta$-decay $Q$ values of $^{130}$Te,
  $^{128}$Te, and $^{120}$Te}, Phys. Rev. C80 (2009) 025501.
\newblock \href {http://arxiv.org/abs/0902.2376} {\path{arXiv:0902.2376}}.

\bibitem{Rahaman_Q_2011}
S.~Rahaman, et~al., {Double-beta decay Q values of 116Cd and 130Te}, Phys.
  Lett. B 703~(4) (2011) 412.

\bibitem{Cuoricino_NDBD_2011}
E.~Andreotti, et~al., {130Te Neutrinoless Double-Beta Decay with CUORICINO},
  Astropart. Phys. 34 (2011) 822--831.
\newblock \href {http://arxiv.org/abs/1012.3266} {\path{arXiv:1012.3266}}.

\bibitem{CUORE_IHE}
D.~R. Artusa, et~al., {The frontier of neutrinoless double beta decay with
  bolometers}, to be submitted (2014).

\bibitem{cryogenic_particle_detection}
C.~Enss (Ed.), Cryogenic Particle Detection, Springer-Verlag Berlin Heidelberg,
  2005.

\bibitem{Arnaboldi:2010fj}
C.~Arnaboldi, et~al., {Production of high purity TeO$_2$ single crystals for
  the study of neutrinoless double beta decay}, J. Cryst. Growth 312~(20)
  (2010) 2999--3008.
\newblock \href {http://arxiv.org/abs/1005.3686} {\path{arXiv:1005.3686}}.

\bibitem{McCammon:2005yj}
D.~McCammon, {Semiconductor thermistors}, Appl. Phys. 99 (2005) 35--61.
\newblock \href {http://arxiv.org/abs/physics/0503086}
  {\path{arXiv:physics/0503086}}.

\bibitem{CUORE0_InitialPerformance_2014}
D.~R. Artusa, et~al., Initial performance of the {CUORE-0} experiment,
  {arXiv:1402.0922}.

\bibitem{TAUP2013_CUORE-0}
D.~R. Artusa, et~al., {First data of CUORE-0}, this proceedings.

\bibitem{CUORE_AHEP_2014}
D.~R. Artusa, et~al., Searching for neutrinoless double-beta decay of
  {$^{130}$Te} with {CUORE}, arXiv:1402.6072.

\bibitem{CUORE_projected_BG}
D.~R. Artusa, et~al., Projected background budget of the {CUORE} experiment, to
  be submitted.

\bibitem{Mei:2005gm}
D.~Mei, A.~Hime, {Muon-Induced Background Study for Underground Laboratories},
  Phys. Rev. D73 (2006) 053004.
\newblock \href {http://arxiv.org/abs/astro-ph/0512125}
  {\path{arXiv:astro-ph/0512125}},

\bibitem{Alessandrello:1998xy}
A.~Alessandrello, et~al., Measurements of internal radioactive contamination in
  samples of {R}oman lead to be used in experiments on rare events, Nucl.
  Instrum. Meth. B142 (1998) 163.

\bibitem{Andreotti:2009dk}
E.~Andreotti, et~al., {Muon-induced backgrounds in the CUORICINO experiment},
  Astropart. Phys. 34 (2010) 18--24.
\newblock \href {http://arxiv.org/abs/0912.3779} {\path{arXiv:0912.3779}},

\bibitem{CUOREExternal}
F.~Bellini, et~al., Monte carlo evaluation of the external gamma, neutron and
  muon induced background sources in the {CUORE} experiment, Astropart.~Phys.
  33 (2010) 169.

\bibitem{GERDA_2013}
M.~Agostini, et~al., {Results on neutrinoless double beta decay of 76Ge from
  GERDA Phase I}, Phys. Rev. Lett. 111 (2013) 122503.
\newblock \href {http://arxiv.org/abs/1307.4720} {\path{arXiv:1307.4720}},

\bibitem{barabash_supernemo_2011}
A.~S. Barabash, {SuperNEMO} double beta decay experiment, {arXiv:1112.1784}.

\bibitem{Clemenza:2011zz}
M.~Clemenza, C.~Maiano, L.~Pattavina, E.~Previtali, {Radon-induced surface
  contaminations in low background experiments}, Eur. Phys. J. C71 (2011) 1805.
\newblock

\bibitem{Cryostat_overview_2012}
A.~Nucciotti, et~al., {Status of the Cryogen-Free Cryogenic System for the
  CUORE Experiment}, Journal of Low Temperature Physics 167 (2012) 528--534.
\newblock

\bibitem{Cryostat_4K_2014}
E.~Ferri, et~al., {Commissioning of the 4 K Outer Cryostat for the CUORE
  Experiment}, Journal of Low Temperature Physics (2014) 1--7

\bibitem{ManciniTerracciano:2012fq}
C.~Mancini-Terracciano, M.~Vignati, {Noise correlation and decorrelation in
  arrays of bolometric detectors}, JINST 7 (2012) P06013.
\newblock \href {http://arxiv.org/abs/1203.1782} {\path{arXiv:1203.1782}}.

\bibitem{TabarellideFatis:2009zz}
T.~Tabarelli~de Fatis, {Cherenkov emission as a positive tag of double beta
  decays in bolometric experiments}, Eur.~Phys.~J. C65 (2010) 359.
\newblock

\bibitem{Beeman:2011yc}
J.~Beeman, et~al., {Discrimination of alpha and beta/gamma interactions in a
  TeO$_2$ bolometer}, Astropart.~Phys. 35 (2012) 558.
\newblock \href {http://arxiv.org/abs/1106.6286} {\path{arXiv:1106.6286}},

\bibitem{Angloher:2011uu}
G.~Angloher, et~al., {Results from 730 kg days of the CRESST-II Dark Matter
  Search}, Eur. Phys. J. C72 (2012) 1971.
\newblock \href {http://arxiv.org/abs/1109.0702} {\path{arXiv:1109.0702}},

\bibitem{Calder_JLTP_2014}
S.~Domizio, et~al., Cryogenic wide-area light detectors for neutrino and dark
  matter searches, J. Low. Temp. Phys. (2014) 1--7

\bibitem{Isaila2012160}
C.~Isaila, et~al., {Low-temperature light detectors: Neganov-Luke amplification
  and calibration}, Phys. Lett. B 716~(1) (2012) 160.
\newblock

\bibitem{Pirr06}
S.~Pirro, et~al., Scintillating double-beta-decay bolometers, Phys. Atom. Nucl.
  69 (2006) 2109.

\bibitem{Arnaboldi:2010jx}
C.~Arnaboldi, et~al., {Characterization of ZnSe scintillating bolometers for
  Double Beta Decay}, Astropart.~Phys. 34 (2011) 344.
\newblock \href {http://arxiv.org/abs/1006.2721} {\path{arXiv:1006.2721}},

\bibitem{Beeman2013JINST}
J.~W. Beeman, et~al., Performances of a large mass {ZnSe} bolometer to search
  for rare events, Journal of Instrumentation 8 (2013) 05021.
\newblock

\bibitem{Arnaboldi:2010tt}
C.~Arnaboldi, et~al., {CdWO$_4$ scintillating bolometer for Double Beta Decay:
  Light and Heat anticorrelation, light yield and quenching factors},
  Astropart.~Phys. 34 (2010) 143.
\newblock \href {http://arxiv.org/abs/1005.1239} {\path{arXiv:1005.1239}},

\bibitem{Beeman2012318}
J.~Beeman, et~al., {A next-generation neutrinoless double beta decay experiment
  based on ZnMoO$_4$ scintillating bolometers}, Phys. Lett. B 710~(2) (2012)
  318.
\newblock

\bibitem{Gironi:2010hs}
L.~Gironi, et~al., {Performance of ZnMoO$_4$ crystal as cryogenic scintillating
  bolometer to search for double beta decay of molybdenum}, JINST 5 (2010)
  P11007.
\newblock \href {http://arxiv.org/abs/1010.0103} {\path{arXiv:1010.0103}},

\bibitem{ZnMoO4_1}
J.~Beeman, et~al., {ZnMoO4: A promising bolometer for neutrinoless double beta
  decay searches}, Astropart.~Phys. 35~(12) (2012) 813.
\newblock

\bibitem{ZnMoO4_2}
J.~Beeman, et~al., {Performances of a large mass ZnMoO$_4$ scintillating
  bolometer for a next generation neutrinoless double beta decay experiment},
  Eur. Phys. J. C72 (2012) 1.

\end{thebibliography}

\end{document}